\documentclass[11pt]{article}
\usepackage{graphicx}

\def\Re{{\cal R \mskip-4mu \lower.1ex \hbox{\it e}\,}}
\def\Im{{\cal I \mskip-5mu \lower.1ex \hbox{\it m}\,}}
\def\ie{{\it i.e.}}
\def\eg{{\it e.g.}}

\def\etal{{\it et al.}}

\def\sub#1{_{\lower.25ex\hbox{$\scriptstyle#1$}}}
\def\tev{\,{\ifmmode\mathrm {TeV}\else TeV\fi}}
\def\gev{\,{\ifmmode\mathrm {GeV}\else GeV\fi}}
\def\mev{\,{\ifmmode\mathrm {MeV}\else MeV\fi}}
\def\mpl{\ifmmode M_{pl}\else $M_{pl}$\fi}
\def\mpl{\ifmmode \overline M_{Pl}\else $\bar M_{Pl}$\fi}
\def\to{\rightarrow}

\def\subw{_{\rm w}}
\def\mh{\ifmmode m\sbl H \else $m\sbl H$\fi}
\def\mch{\ifmmode m_{H^\pm} \else $m_{H^\pm}$\fi}
\def\mt{\ifmmode m_t\else $m_t$\fi}
\def\mc{\ifmmode m_c\else $m_c$\fi}
\def\mz{\ifmmode M_Z\else $M_Z$\fi}
\def\mw{\ifmmode M_W\else $M_W$\fi}
\def\mws{\ifmmode M_W^2 \else $M_W^2$\fi}
\def\mhs{\ifmmode m_H^2 \else $m_H^2$\fi}   
\def\mzs{\ifmmode M_Z^2 \else $M_Z^2$\fi}
\def\mts{\ifmmode m_t^2 \else $m_t^2$\fi}
\def\mcs{\ifmmode m_c^2 \else $m_c^2$\fi}
\def\mchs{\ifmmode m_{H^\pm}^2 \else $m_{H^\pm}^2$\fi}
\def\ztwo{\ifmmode Z_2\else $Z_2$\fi}
\def\zone{\ifmmode Z_1\else $Z_1$\fi}
\def\mtwo{\ifmmode M_2\else $M_2$\fi}
\def\mone{\ifmmode M_1\else $M_1$\fi}
\def\tb{\ifmmode \tan\beta \else $\tan\beta$\fi}
\def\xw{\ifmmode x\subw\else $x\subw$\fi}
\def\ch{\ifmmode H^\pm \else $H^\pm$\fi}
\def\lum{\ifmmode {\cal L}\else ${\cal L}$\fi}
\def\inpb{\,{\ifmmode {\mathrm {pb}}^{-1}\else ${\mathrm {pb}}^{-1}$\fi}}
\def\infb{\,{\ifmmode {\mathrm {fb}}^{-1}\else ${\mathrm {fb}}^{-1}$\fi}}
\def\epem{\ifmmode e^+e^-\else $e^+e^-$\fi}
\def\ppb{\ifmmode \bar pp\else $\bar pp$\fi}
\def\bsg{\ifmmode B\to X_s\gamma\else $B\to X_s\gamma$\fi}
\def\bsll{\ifmmode B\to X_s\ell^+\ell^-\else $B\to X_s\ell^+\ell^-$\fi}
\def\bstt{\ifmmode B\to X_s\tau^+\tau^-\else $B\to X_s\tau^+\tau^-$\fi}
\def\lamt{\ifmmode \tilde\lambda\else $\tilde\lambda$\fi}
\def\shat{\ifmmode \hat s\else $\hat s$\fi}
\def\that{\ifmmode \hat t\else $\hat t$\fi}
\def\uhat{\ifmmode \hat u\else $\hat u$\fi}

\newskip\zatskip \zatskip=0pt plus0pt minus0pt
\def\matth{\mathsurround=0pt}
\def\lsim{\mathrel{\mathpalette\atversim<}}
\def\gsim{\mathrel{\mathpalette\atversim>}}
\def\atversim#1#2{\lower0.7ex\vbox{\baselineskip\zatskip\lineskip\zatskip
  \lineskiplimit 0pt\ialign{$\matth#1\hfil##\hfil$\crcr#2\crcr\sim\crcr}}}

\def\grtsim{\,\,\rlap{\raise 3pt\hbox{$>$}}{\lower 3pt\hbox{$\sim$}}\,\,}
\def\lsim{\,\,\rlap{\raise 3pt\hbox{$<$}}{\lower 3pt\hbox{$\sim$}}\,\,}

\renewcommand{\thefootnote}{\fnsymbol{footnote}}

\begin{document} \begin{titlepage}
\rightline{\vbox{\halign{&#\hfil\cr
&SLAC-PUB-11599\cr
}}}
\begin{center}
\thispagestyle{empty} \flushbottom { {\Large\bf TeV-Scale Black Hole Lifetimes 
in Extra-Dimensional Lovelock Gravity 
\footnote{Work supported in part
by the Department of Energy, Contract DE-AC02-76SF00515}
\footnote{e-mail:
$^a$rizzo@slac.stanford.edu}}}
\medskip
\end{center}

\centerline{Thomas G. Rizzo$^{a}$}
\vspace{8pt} 
\centerline{\it Stanford Linear
Accelerator Center, 2575 Sand Hill Rd., Menlo Park, CA, 94025}

\vspace*{0.3cm}

\begin{abstract}
We examine the mass loss rates and lifetimes of TeV-scale extra dimensional black holes (BH) 
in ADD-like models with Lovelock higher-curvature terms present in the action. In particular 
we focus on the predicted differences between the canonical and microcanonical ensemble  
statistical mechanics descriptions of the Hawking radiation that results in the decay of 
these BH. In even numbers of extra dimensions the employment of the microcanonical approach 
is shown to generally lead to a significant increase in the BH lifetime as in case of the 
Einstein-Hilbert action. For odd numbers of extra dimensions, stable BH remnants occur 
when employing either description provided the highest order allowed Lovelock invariant is 
present. However, in this case, the time dependence of the mass loss rates obtained employing 
the two approaches will be different. These effects are in principle measurable at 
future colliders. 
\end{abstract}



\renewcommand{\thefootnote}{\arabic{footnote}} \end{titlepage} 

%
%
%

\section{Introduction and Background}

The large extra dimensions picture of Arkani-Hamed, Dimopoulos and 
Dvali(ADD){\cite {ADD}} suggests that the fundamental 
scale of gravity, $M_*$, may not be far above the weak scale $\sim$ TeV. 
In this scenario, gravity propagates in the $D=4+n$ dimensional bulk  
while the Standard Model(SM) is confined to a 
three-dimensional 'brane' which is assumed to be flat. In such a scenario 
one finds that $M_*$ 
is related to the usual 4-d (reduced) Planck scale, $\mpl$, via the expression 
\begin{equation}
\mpl^2=V_nM_*^{n+2}\,,
\end{equation}
where $V_n$ is the volume of the compactified extra dimensions. Assuming 
for simplicity that they form an $n$-dimensional torus, if all compactification 
radii, $R_c$, are the same, then $V_n=(2\pi R_c)^n$. This basic  
ADD picture leads to three essential predictions {\cite {JM}}: ($i$) the 
emission of graviton Kaluza-Klein(KK) states during the collision of SM particles 
leading to signatures with apparent missing energy{\cite {GRW,HLZ,PP}}; ($ii$) 
the exchange of graviton KK excitations between SM fields leading to dimension-8 contact 
interaction-like operators with distinctive spin-2 properties{\cite {GRW,HLZ,JLH}; 
($iii$) the production of black holes(BH) at colliders and in cosmic rays 
with geometric cross sections, $\simeq \pi R_s^2$, with $R_s$ being the BH 
Schwarzschild radius, once collision   
energies greater than $\sim M_*$ are exceeded{\cite {Fisch,DL,GT,Kanti}}. 
{\footnote {Note that in the 
simplest picture the BH production threshold is just a simple step-function.}}   
It has been noted that while ($i$) and ($ii$) are the result of 
an expansion of the $D-$dimensional 
Einstein-Hilbert(EH) action to leading order in the gravitational field and 
are in some sense perturbative, ($iii$) on the other-hand relies upon the full 
non-perturbative content of the EH action. Thus TeV scale BH production is actually 
testing $D-$dimensional General Relativity and not {\it just} the ADD picture. This is 
important as many other alternative theories of gravity in extra dimensions can lead to 
the same leading order graviton interactions.  
Within the ADD scenario, future collider measurements of the ($i$) and ($ii$) type 
processes should be able to tell us the values of both the quantities 
$n$ and $M_*${\cite {JM}} rather precisely. 

Of course ADD 
is at best an effective theory that operates at energies below the scale $M_*$. 
It is reasonable to expect that at least some aspects of the full UV theory 
may leak down into these collider tests and may lead to potentially 
significant quantitative and/or qualitative modifications of simple ADD expectations 
that can be probed experimentally. 
We have recently begun an examination of the effect of the presence of 
higher curvature invariants in the $D-$dimensional action 
of ADD-like models {\cite {Rizzo:2005fz}} as well as in models with a warped
metric{\cite {RS}}. We note that since the ADD bulk
is flat and the SM fields are confined to a brane, the predictions for
($i$) and ($ii$) above are not influenced by the addition of such extra terms in
the action{\cite {{Demir:2005ps}} as the analogous predictions would be in the 
case of warped extra dimensions. Motivated by
string theory{\cite {Zwiebach,big,Mavromatos}}, we examined a very special class of
such invariants with interesting properties first described by
Lovelock{\cite {Lovelock}}, called Lovelock invariants. 

Lovelock invariants come in fixed order, $m$, which we denote as ${\cal L}_m$,
that describes the number of powers of the curvature tensor out of which they
are constructed. We can express the ${\cal L}_m$ as
\begin{equation}
{\cal L}_m \sim \delta^{A_1B_1...A_mB_m}_{C_1D_1...C_mD_m}~R_{A_1B_1}
~^{C_1D_1}.....R_{A_mB_m}~^{C_mD_m}\,,
\end{equation}
where $\delta^{A_1B_1...A_mB_m}_{C_1D_1...C_mD_m}$ is the totally
antisymmetric product of Kronecker deltas and $R_{AB}~^{CD}$ is the
$D$-dimensional curvature tensor.
For a space with an even number of dimensions, the $D=2m$
Lovelock invariant is topological and leads to a total derivative, \ie,
a surface term, in the action.  All of the higher order invariants,
$D\leq 2m-1$, can then be shown to vanish identically by using curvature
tensor index symmetry properties. On the otherhand, for the cases with
$D\geq 2m+1$, the ${\cal L}_m$ are truly dynamical objects that once 
added the action can significantly alter the field equations usually
associated with the EH term. However, it can be shown that the addition of
any or all of the ${\cal L}_m$ to the EH action still results in a theory
with only second order equations of motion as in
ordinary Einstein gravity. In particular, variation of the action
leads to modifications of Einstein's equations by the addition of
new terms which are second-rank symmetric tensors with vanishing covariant
derivatives, depending only on the metric and its first and second
derivatives, \ie, they have the same general properties as the Einstein
tensor itself but are higher order in the curvature. These are quite
special properties not possessed by arbitrary invariant structures which usually 
lead to equations of motion of higher order, \ie, more co-ordinate
derivatives of the metric tensor and graviton field, \eg, terms with quartic
derivatives. Such theories would, in general, have serious problems with the
presence of tachyons and ghosts as well as with perturbative 
unitarity{\cite {Zwiebach}}. The Lovelock invariants are constructed in such a way 
as to produce an action which is free of these problems. In addition, as might be 
expected, the introduction of Lovelock terms into the action does not modify the 
number of degrees of freedom encountered by studying the EH action. {\footnote {As 
is well known, the addition of {\it arbitrary} curvature invariants to the EH action 
can lead to new propagating degrees of freedom in the resulting equations of motion.}}

In our earlier work{\cite {Rizzo:2005fz}} we showed that the presence in the action
of Lovelock invariants can lead to TeV-scale BH in 
ADD-like models with thermodynamical properties 
that can significantly differ from the usual EH expectations. This includes the
possibility that BH may be stable in $n$-odd dimensions and that have production
cross sections with calculable mass thresholds. In a more general context, BH in 
theories with Lovelock invariants have been
discussed by a large number of authors{\cite {big}}. The usual thermodynamical 
description of the Hawking radiation produced by TeV-scale BH decays is via the 
canonical ensemble(CE){\cite {Kanti}} which has been employed in most analyses 
in the literature (in particular, our previous analysis of ADD-like BH).
However, as pointed out by several groups{\cite {Casadio:2001dc}}, 
though certainly applicable to very massive BH, this approach 
does not strictly apply when $M_{BH}/M_*$ is not much greater than O(1) or when the 
emitted particles carry an energy comparable to the BH mass itself due to 
the back-reaction of the emitted particles on the properties of the BH. This certainly 
happens when the resulting overall BH Hawking radiation multiplicity is low.  
In the decay of TeV-scale BH that can be made at a collider, 
the energy of the emitted particles is generally comparable to both $M_*$ as well as 
the mass of the BH itself thus requiring the MCE treatment. 
In the CE approach the BH is treated as a large 
heat bath whose temperature is not significantly  
influenced by the emission of an individual particle. While this is a very good  
approximation for reasonably heavy BH it becomes worse as the BH mass approaches the 
$M_*$ scale as it does for the case we consider below. 
Furthermore, the BH in an asymptotically flat space (which we can assume 
here since the BH Schwarzschild radii, $R_s$, are far 
smaller than $R_c$) cannot be in equilibrium with its Hawking radiation.  

It has been suggested{\cite {Casadio:2001dc}} 
that all these issues can be dealt with simultaneously 
if we instead employ the correct, \ie, microcanonical ensemble(MCE) approach 
in the statistical mechanics treatment for BH 
decay. As $M_{BH}/M_*$ grows larger, $\gsim 10-20$, the predictions of these two 
treatments will be found to agree, but they differ in the region which is of most  
interest to us since at colliders 
we are close to the BH production threshold where $M_{BH}/M_*$ 
is not far above unity. Within the framework of the 
EH action it has been emphasized{\cite {Casadio:2001dc}} 
that TeV-scale BH lifetimes will be increased by many orders of magnitude when the MCE 
approach is employed in comparison to the conventional CE expectations. This is not due 
to modifications in the thermodynamical quantities, such as the temperature, themselves 
but how they enter the expressions for the rate of mass loss in the decay of the BH. 
Here we will address the issue of how these two statistical 
descriptions may differ in the BH mass range of interest to us when the additional higher-curvature 
Lovelock terms are present in the action. In particular, we need to address 
what the combination of Lovelock terms plus the MCE description do to the BH mass loss 
rates and lifetimes. In the case of $n$ even we will show that BH lifetimes are 
significantly increased as was found in the EH case. 
In the case of odd $n$ the Lovelock BH will of course be found to produce  
stable remnants using either prescription for the same set of parameters as we will see below.

The outline of the paper is as follows: 
In Section 2 we will present the basic ingredients associated with the Lovelock 
invariant extended action and the altered expectations for the Schwarzschild 
radius, temperature and entropy of TeV-scale BH in ADD-like models where the bulk 
is essentially flat. We then will provide a brief overview of the general 
formalism for calculating BH lifetimes using the MCE, contrasting with the conventional 
CE approach. In Section 3 we will perform a numerical comparison of the predictions 
for the BH mass loss rate and lifetime in both the CE and MCE frameworks. For 
purposes of comparison we first analyze the results when only the EH term is present 
in the action. We also address the issue as to whether the dominance of BH decays to brane 
fields is influenced by the choice of thermodynamic description. Section 4 contains a 
discussion and our conclusions.

\section{Formalism}

Based on the discussion above and the definition of the ${\cal L}_{m}$ we see that 
the most general action with Lovelock invariants in 4-d is just EH plus a 
cosmological constant, \ie, ordinary General Relativity. In 5-d, all of the 
${\cal L}_{m \geq 3}$ still vanish as in 4-d but ${\cal L}_{2}$, which 
is the familiar Gauss-Bonnet(GB) invariant, 
is no longer a total derivative and its presence will modify the results 
obtained from Einstein gravity. The generalization is clear: 
for $D=5,6$ only ${\cal L}_{0-2}$ can be present in the action. For $D=7,8$ only 
${\cal L}_{0-3}$ can be present while for $D=9,10$ only ${\cal L}_{0-4}$. 
Since ADD assumes that the compactified space is {\it flat} the coefficient of 
${\cal L}_{0}$ is taken to be zero in the present framework 
and, to reproduce the correct limit,the coefficient of   
${\cal L}_{1}$ is normalized so that it can be identified with the usual EH term. 
Thus in the Lovelock-ADD picture for $D \leq 10(n \leq 6)$  
there are at most three new pieces to add to the EH action and so  
the general form relevant for the extended ADD model we consider is given by 
\begin{equation}
S={M_*^{n+2}\over {2}}\int d^{4+n}x ~\sqrt {-g}~\Bigg[R+
{\alpha\over {M_*^2}}{\cal L}_2+{\beta \over {M_*^4}}{\cal L}_3+
{\gamma \over {M_*^6}}{\cal L}_4\Bigg]\,,
\end{equation}
where $\alpha$, $\beta$ and $\gamma$ are dimensionless coefficients which we 
take to be {\it positive} in the discussion below. (If we consider $D>10$ it is 
quite easy to extend this parameterization by including potential ${\cal L}_5$ 
contributions.) If we expect this expansion to 
be the result of some sort of perturbation theory,    
some algebra suggests that $\alpha D^2,\beta D^4$, and 
$\gamma D^6 \leq 1$ which yields the (only) suggestive values $\alpha \sim 10^{-2}$, 
$\beta \sim 10^{-3}-10^{-4}$ and $\gamma \sim 10^{-5}$. 
{\footnote {It is important to note that the 
fundamental mass parameter, $M_*$ appearing in the above action is the 
{\it same} as the one appearing in the ADD $M_*$-$\mpl$ relationship Eq.(1). It is 
also the parameter appearing in the 5-d coupling of the graviton to matter fields.  
$M_*$ can be directly related to several other mass parameters used in the 
literature. The fundamental scale employed by Dimopoulos and Landsberg{\cite {DL}} is 
given by $M_{DL}=(8\pi)^{1/(n+2)}M_*$ while that of Giddings and 
Thomas{\cite {GT}} is found to be $M_{GT}=[2(2\pi)^n]^{1/(n+2)}M_*$;  
moreover, Giudice \etal {\cite {GRW}} employ a different scale,  
$M_{GRW}=(2\pi)^{n/(n+2)}M_*$. $M_*$ is thus correspondingly 
{\it smaller} than all of these other parameters with consequently far weaker 
experimental bounds{\cite {JM}}. For example, if 
$n=2(6)$ and $M_{GRW}>1.5$ TeV then $M_*>0.60(0.38)$ TeV; existing direct bounds on 
$M_*$ are thus well below 1 TeV at present.}}

In order to calculate BH lifetimes we will need the relationships between the BH 
mass, the Schwarzschild radius, $R_s$, temperature, $T$ and entropy, $S$. When 
Lovelock terms are present in the action these relations can be significantly 
modified from their conventional EH expectations. Here we simply summarize results  
from our earlier work. (For details, see 
{\cite {Rizzo:2005fz}}). The $M_{BH}-R_s$ relationship is given by
\begin{eqnarray}
m(x)&=&c\Big[x^{n+1}+\alpha n(n+1)x^{n-1}+\beta n(n+1)(n-1)(n-2)x^{n-3}
\nonumber \\
 &+&\gamma n(n+1)(n-1)(n-2)(n-3)(n-4)x^{n-5}\Big]\,,
\end{eqnarray}
where $m=M_{BH}/M_*$, $x=M_*R_s$, and the numerical constant $c$ is given by
\begin{equation}
c={{(n+2) \pi^{(n+3)/2}}\over {\Gamma({{n+3}\over {2}})}}\,.
\end{equation}
Since what we really want to know is $x(m)$ and not $m(x)$ as given above we 
must find the roots of this polynomial equation; this must be done 
numerically in general except for some special cases. Fortunately, for 
the range of parameters of interest to us and with $\alpha,\beta, \gamma\geq 0$, 
we find that this polynomial has only one distinct real positive root. 
Similarly, the BH temperature is found to be given by 
\begin{equation}
T={(n+1)\over {4\pi}}{U(x)\over {V(x)}}\,,
\end{equation}
where $T=T_{BH}/M_*$ and 
\begin{eqnarray}
U(x)&=&x^6+\alpha n(n-1)x^4+\beta n(n-1)(n-2)(n-3)x^2\nonumber \\
    &+&\gamma n(n-1)(n-2)(n-3)(n-4)(n-5)\nonumber \\
V(x)&=&x\Big[x^6+2\alpha n(n+1)x^4+3\beta n(n+1)(n-1)(n-2)x^2\nonumber \\
    &+&4\gamma n(n+1)(n-1)(n-2)(n-3)(n-4)\Big]\,. 
\end{eqnarray}
The corresponding BH 
entropy can then be calculated using the familiar thermodynamical relation 
\begin{equation}
S=\int_0^x dx ~T^{-1} {\partial m\over {\partial x}}\,,
\end{equation}
which yields 
\begin{eqnarray}
S&=&{{4\pi c}\over {n+2}}\Big[x^{n+2}+2\alpha (n+1)(n+2)x^n+3\beta n(n+1)(n+2)
(n-1)x^{n-2}\nonumber \\
&+&4\gamma n(n+1)(n+2)(n-1)(n-2)(n-3)x^{n-4}\Big]\,. 
\end{eqnarray}
Note that here we have required that the entropy vanish for a zero horizon size. 

In order to calculate the BH mass loss rates we follow the formalism in  
{\cite {Casadio:2001dc}}; to simplify our presentation and to focus 
on the differences between thermodynamical treatments we will {\it ignore} the 
effects due to grey-body factors{\cite {Kanti}} in the present analysis. 
{\footnote {It is interesting to note 
that the presence of Lovelock invariants in the action can alter the usual 
values obtained for scalar, fermion and gauge 
grey-body factors by terms of order unity. However, for the 
range of parameters of interest to us it has recently 
been shown that $\alpha \neq 0$ does not lead to any significant change in 
the bulk or brane grey-body factors from those obtained from 
EH{\cite{Kanti2}}. Our expectation is that similar results will hold when 
$\beta,\gamma \neq 0$ contributions are included but this needs to be explored 
directly. These results need, however, to be fully completed for the graviton modes.}} 
In this approximation, the rate 
of BH mass loss (time here being measured in units of $M_*^{-1}$) into {\it bulk} fields 
employing the MCE approach is given by 
\begin{equation}
\Big[{{dm}\over {dt}}\Big]_{bulk}=
{{\Omega_{d+3}^2}\over {(2\pi)^{d+3}}} ~\zeta(d+4) x^{d+2}
\sum_i ~N_i\int^m_{m_{crit}} ~dy ~(m-y)^{d+3}\Big [e^{S(m)-S(y)}+s_i\Big]^{-1}\,,
\end{equation}
where, $x=M_*R_s$ as above, $S$ in the corresponding 
entropy of the BH, $d$ labels the number 
of bulk dimensions into which the particles are emitted, $d \leq n$, 
$i$ labels various particle species with $N_i$ degrees of freedom 
which live in the bulk and obey 
Fermi-Dirac(Boltzmann, Bose-Einstein) statistics, corresponding to the choices of 
$s_i=1(0,-1)$, $\zeta$ is the familiar Riemann Zeta Function and as usual 
\begin{equation}
\Omega_{d+3}={{2\pi^{(d+3)/2}}\over {\Gamma((d+3)/2))}}\,.
\end{equation}
The lower limit of the integration 
$m_{crit}$=0 for $n$-even but may be finite for $n$-odd corresponding to the mass 
threshold for BH production which can occur in the case of the Lovelock extended 
action{\cite {Rizzo:2005fz}}. The value of this parameter can in all cases be 
obtained by first fixing the value of $n$ and then setting $x=0$ on the right-hand side 
of the expression for $m(x)$ in 
Eq.(4) above. In the present analysis, we will assume that only gravitons live in the 
bulk, so for them $d=n$. For the case of decays into fields that live on the brane, 
the expression is the same as that given above but now with $d=0$ and we now must sum over 
all particles that live only on the brane, \ie, those of the SM. 

\begin{figure}[htbp]
\centerline{
\includegraphics[width=8.5cm,angle=90]{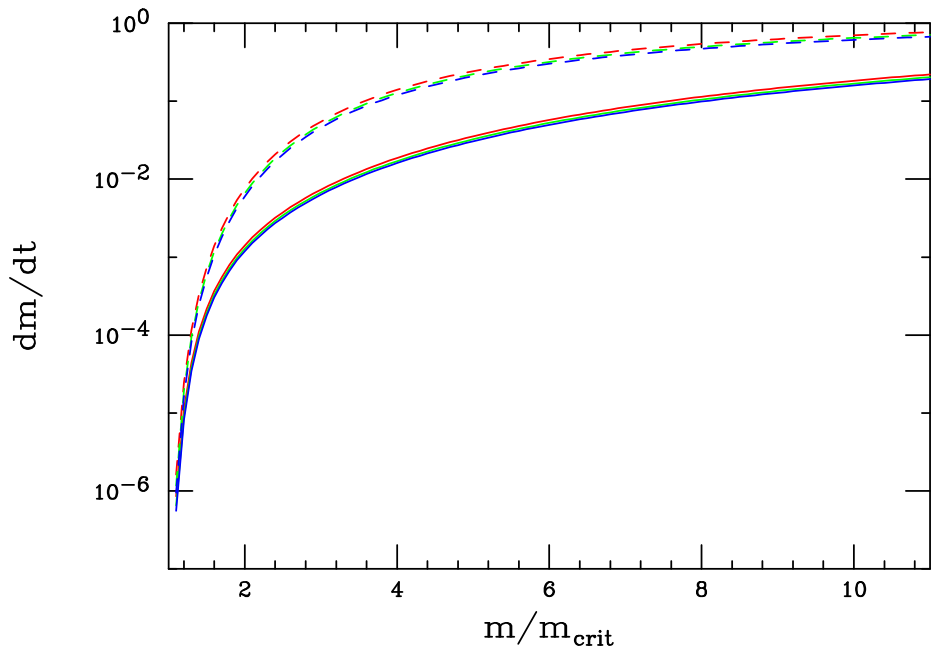}}
\vspace*{0.1cm}
\caption{Comparison of the BH mass loss rates to brane fields following the 
MCE prescription assuming final 
states which, from top to bottom in each set of curves, are purely Fermi-Dirac, Boltzmann 
or Bose-Einstein. The solid(dashed) set corresponds to $n=5, \alpha=0.005, \beta=0.0003, 
\gamma=1.14 \times 10^{-5}$$(n=3, \alpha=0.01, \beta=005$).}
\label{fig0}
\end{figure}

In our recent analysis{\cite {Rizzo:2005fz}} of BH decay in the RS model where GB terms 
were present, we noted that 
following the MCE approach there is some potential sensitivity of the mass loss rate to the 
specific statistics mix of particles on the brane into which the BH decays. 
However, there it was shown that since we are concerned with decays to SM brane 
fields, where the numbers of fermionic degrees of freedom (48, assuming only light 
Dirac neutrinos) is somewhat larger than the number for bosons (14), the value of $s_i=0$ 
was found to be a reasonable approximation given the other uncertainties in the 
calculation. The reason for this is that 
($i$) the SM is mostly fermionic and the results for Boltzmann and FD statistics are   
found to lie numerically 
rather close to one another and ($ii$) the B distribution lies between 
the BE and FD ones. For the cases we study below it is important to check what happens 
in greater numbers of flat dimensions when 
additional Lovelock terms are present; Fig,~\ref{fig0} shows the results of two sample 
calculations. Here we see that for a flat bulk with larger values of $n$ 
and when the new Lovelock terms are present 
with positive values of the coefficients $\alpha,\beta$,and $\gamma$, the 
differences between the predictions of the different statistics final states is 
significantly reduced. The corresponding integrated lifetimes are also found to agree 
within $\sim 10\%$. Thus in 
our numerical analysis that follows we will for simplicity take $s_i=0$ and assume that 
the total number of SM degrees of freedom is $60$.

In the CE approach, the expression above, Eq.(10), simplifies significantly as in this 
case the integral can be performed analytically. The reason for this 
is that in the CE treatment, the factors $S(m)$ and $S(y=m-\omega)$, $\omega$ being 
the scaled 
energy of the blackbody radiation, appearing in the exponential factor above 
are considered nearly the same since backreaction is neglected, \ie, 
taking in the MCE expression above $m\to \infty$ (the no recoil limit) and 
$S(m)-S(m-\omega)\simeq \omega\partial_m S=\omega/T$, then integrating over 
$\omega$ yields the usual CE result. 
In particular, allowing for the different possible particle statistics 
in the CE case as well we obtain for bulk decay
\begin{equation}
\Big[{{dm}\over {dt}}\Big]_{bulk}=\sum_i N_i Q_i 
{{\Omega_{d+3}^2}\over {(2\pi)^{d+3}}} ~\zeta(d+4)x^{d+2} \Gamma(d+4)T^{d+4}\,,
\end{equation}
where $N_i$ is again just the appropriate number of degrees of freedom for each 
statistics type and 
$Q_i$ takes the value $\pi^4/90(1,~7\pi^4/720)$ for BE(B, FD) statistics. 
The corresponding expression for brane decays in the CE case is straightforwardly 
obtained by  setting $d=0$. Note that the values of $Q_i$ differ from each other by less 
than $\sim 10\%$ resulting in only small differences in lifetimes. 
It is interesting to note that since the Hawking radiation emitted from BH is generally 
softer in the MCE approach in comparison to the CE one, the average multiplicity for a fixed 
initial BH mass and number of dimensions is found to be somewhat larger in the MCE case. This 
difference should be observable at future colliders.

Given the large number of SM brane fields, it is well known that for the EH action  
the brane modes tend to dominate over bulk modes by a factor of order $\sim 100$ or more, 
a result which seems to continue to hold when GB terms (and the  
corresponding brane field grey-body factors for scalar, fermion, and gauge emissions) 
are included{\cite {Kanti2}}. It is to be noted that these results were obtained without the inclusion 
of grey-body factors for bulk graviton emission. The first question to 
address is whether these results remain valid when further Lovelock terms are present 
in the action. As we will see in the next section, in the absence of grey-body factors, 
while the bulk to brane ratio increases with $D$ it never exceeds unity when Lovelock terms are 
present in either MCE or CE descriptions{\cite {comment}}. A full analysis including all grey-body 
effects is clearly needed.

\begin{figure}[htbp]
\centerline{
\includegraphics[width=8.5cm,angle=90]{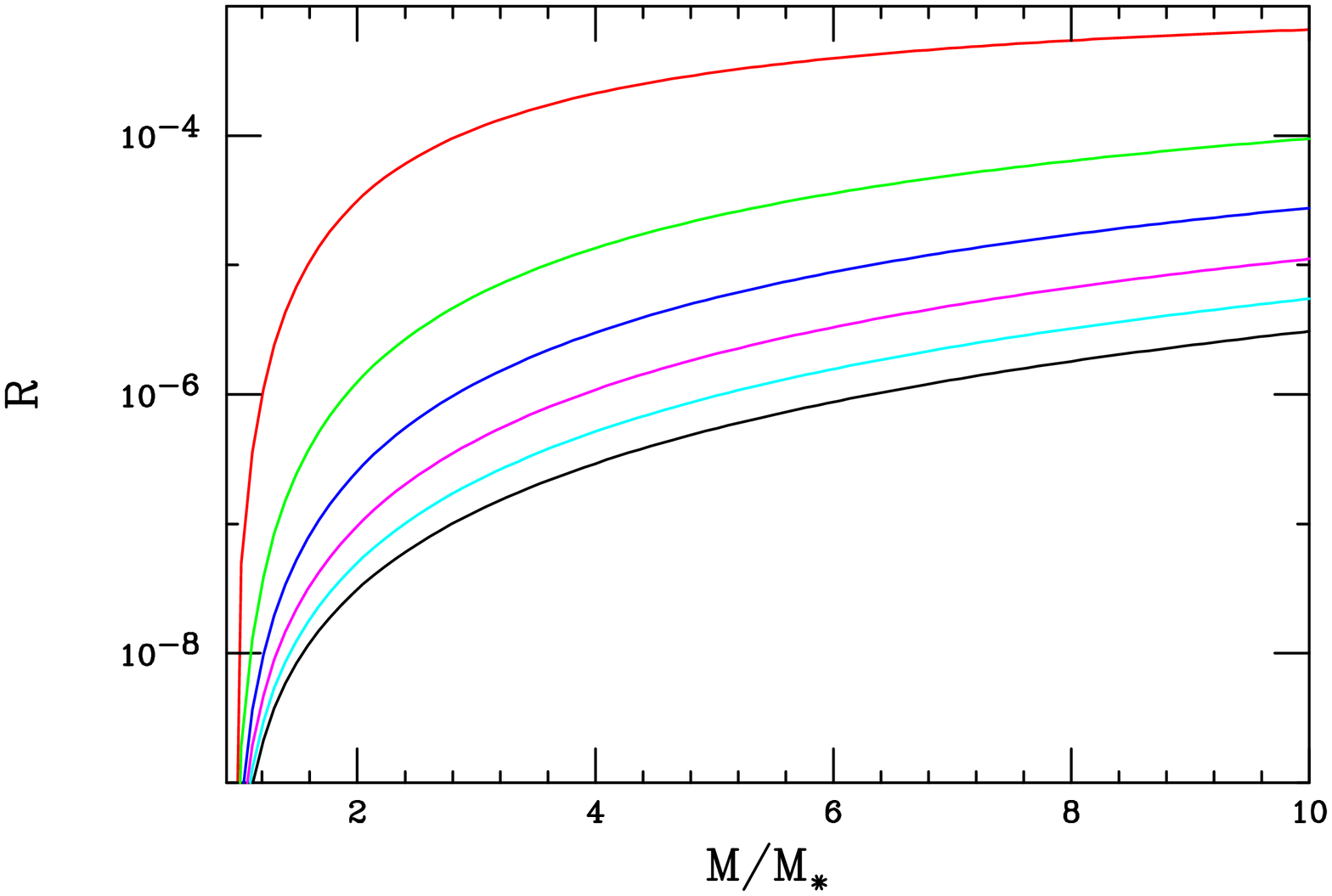}}
\vspace{0.4cm}
\centerline{
\includegraphics[width=8.5cm,angle=90]{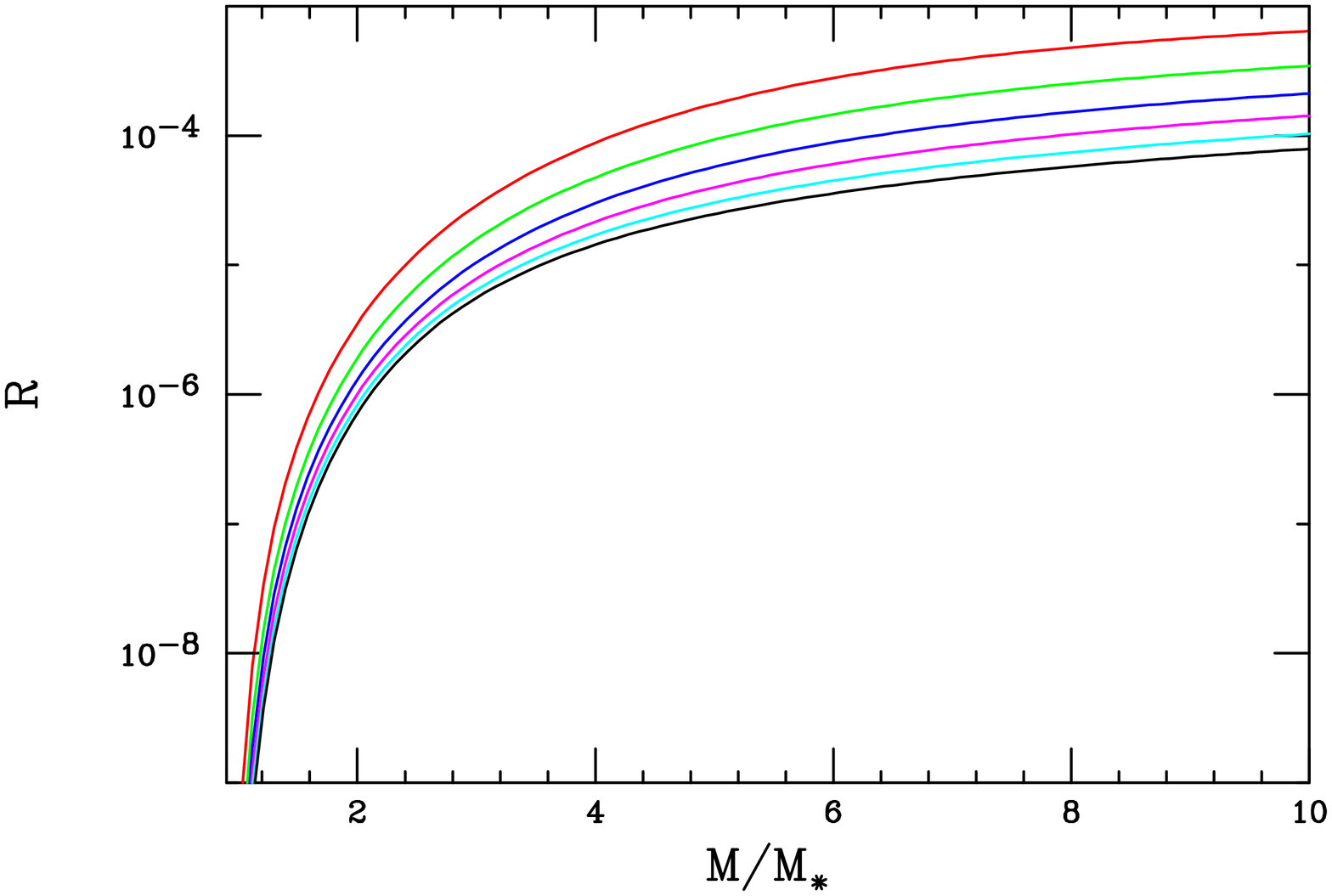}}
\vspace*{0.1cm}
\caption{The ratio $R$ as a function of $m=M_{BH}/M_*$ assuming $n=3$ for $\alpha=0$ to 
$\alpha=0.025$ from top to bottom in steps of 0.005 assuming $\beta=0.0005$. 
The top(bottom) panel is the CE(MCE) result.}
\label{fig1}
\end{figure}

\section{Analysis}

In order to address the above question of how the MCE vs CE choice might 
influence the ratio of bulk to brane BH mass loss rates in the absence of grey-body factors, 
we construct the ratio 
\begin{equation}
R=\Big[{{dm}\over {dt}}\Big]_{bulk}/\Big[{{dm}\over {dt}}\Big]_{brane}\,. 
\end{equation}
Using $s_i=0$, we will assume $N_{brane}=60$ and $N_{bulk}=1$ in what follows and thus 
we might naively expect that $R \sim 10^{-3}-10^{-2}$ if EH provides a reasonable 
estimate over most of the parameter range. Does this estimate remain valid when 
Lovelock terms are present and does the result depend on whether one chooses 
to follow the MCE or CE approach? To be specific in addressing these issues, 
we consider the cases of $n=3,5$ and examine the ratio $R$ as a function 
of $m$ for different values of the Lovelock parameters. The results of this 
analysis are shown in Figs.~\ref{fig1} and ~\ref{fig2} where we see that $R$ is a 
strong and increasing function of $m$ and that $R$ also increases with $n$. 
For large $m$, indeed $R \sim 10^{-3}$ but the precise value depends in 
detail upon the value of 
$n$ as well as the Lovelock parameters in a rather sensitive fashion. As the BH 
decays and $m$ becomes smaller, $R$ rapidly goes to zero in all cases implying a very strong 
suppression of the bulk modes. While the 
CE and MCE approach results do differ in detail they are qualitatively similar and in no 
cases do we see any significant BH decay into bulk states in our mass range of interest.  
(Note, however, that $R$ is generally larger in the MCE case.) Thus we can concentrate 
on brane modes in what follows ignoring the decays to bulk gravitons. Even if grey-body 
factors modify the values of $R$ obtained above 
by an order of magnitude or more when Lovelock terms are 
present we expect the basic result of dominant brane field BH mass loss to remain valid. 

\begin{figure}[htbp]
\centerline{
\includegraphics[width=8.5cm,angle=90]{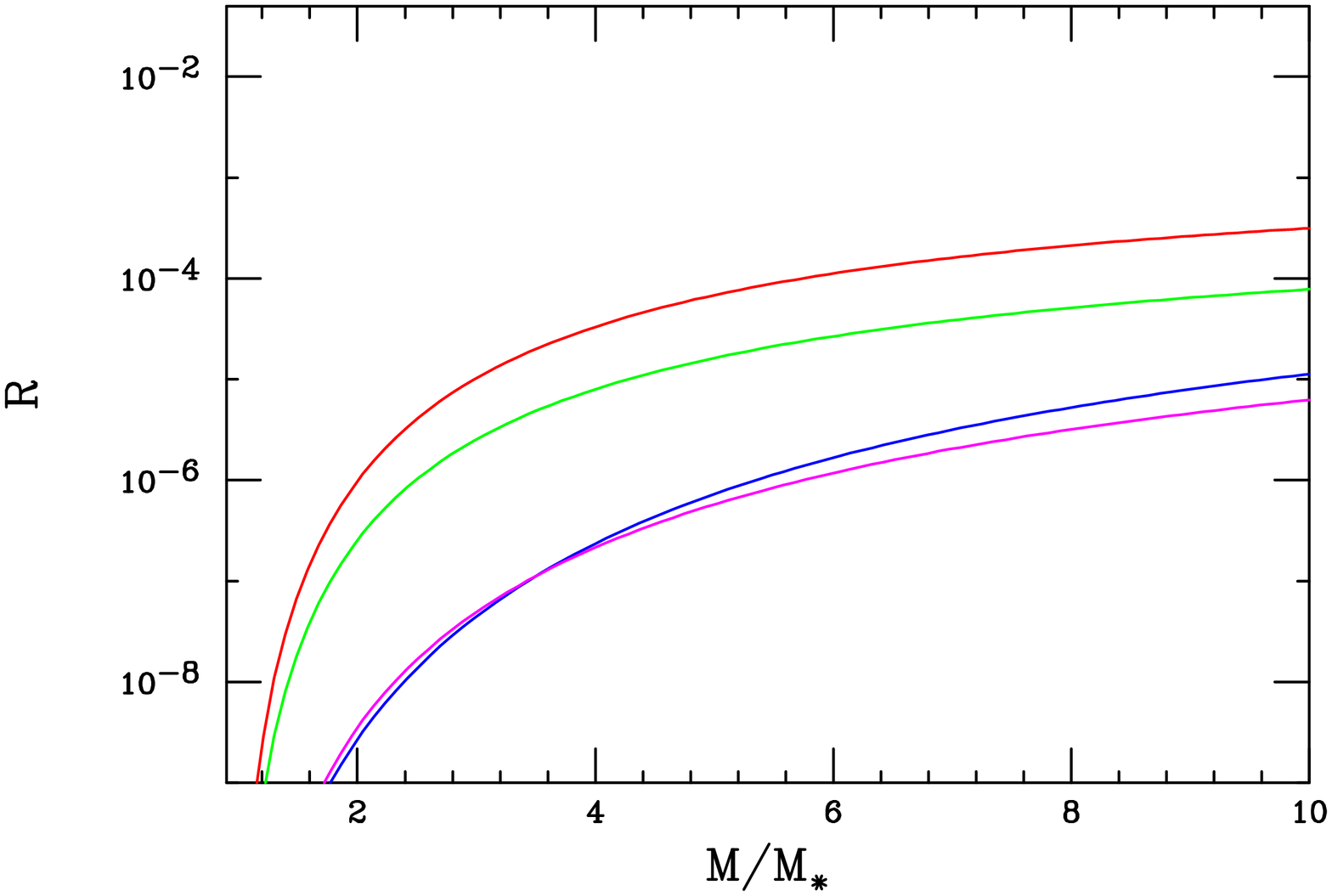}}
\vspace{0.4cm}
\centerline{
\includegraphics[width=8.5cm,angle=90]{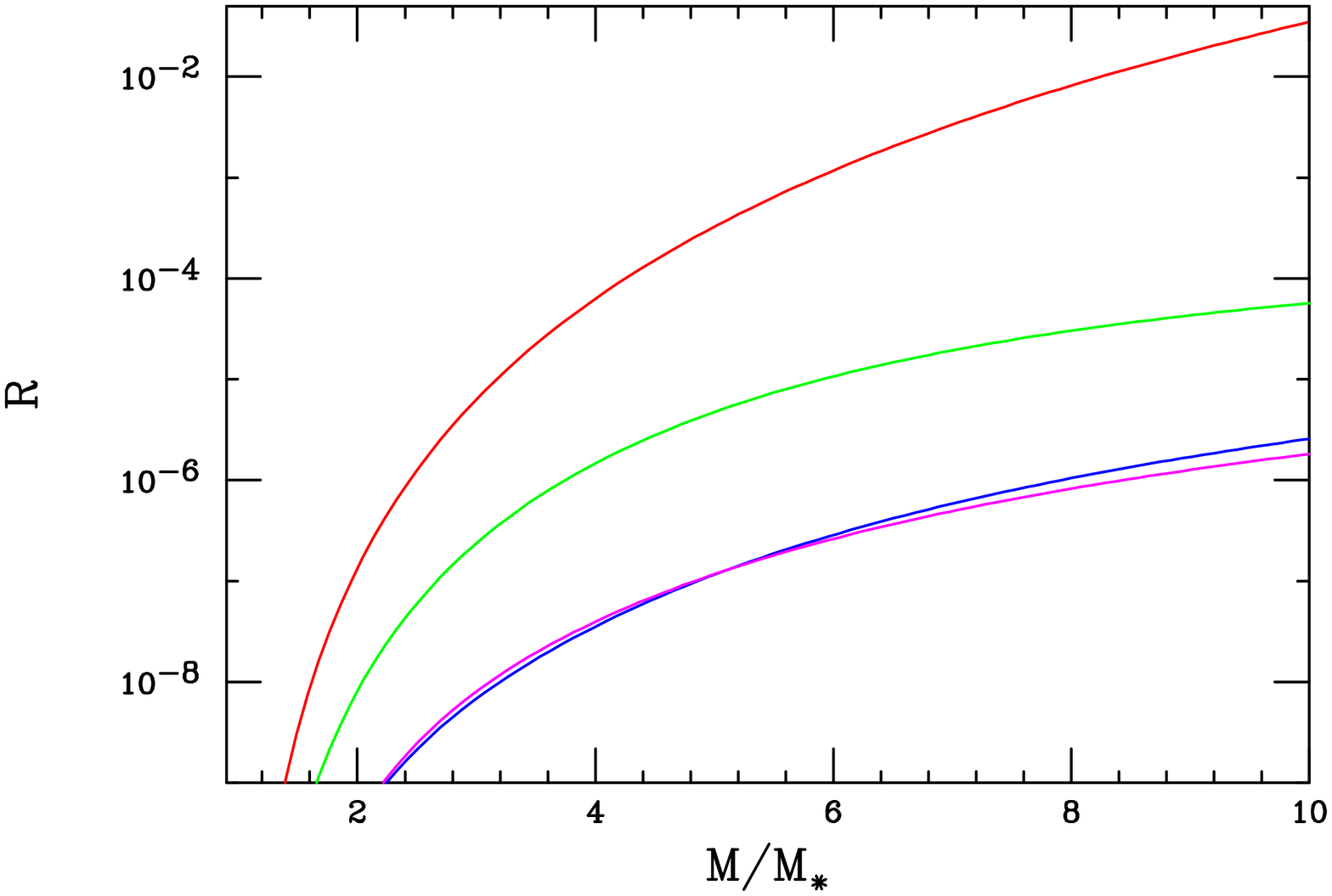}}
\vspace*{0.1cm}
\caption{Same as in the previous figure but now for $n=5$ assuming 
$\gamma=1.14\times 10^{-5}$. From top to bottom on the right-hand side the figure the  
curves correspond to $\alpha=\beta=0$, $\alpha=0.005$ with $\beta=$, $\beta=0.0003$ with 
$\alpha=0$ and $\alpha=0.005,\beta=0.0003$, respectively.}
\label{fig2}
\end{figure}

Before addressing the more complex case of the extended action with Lovelock terms 
let us  briefly review the 
numerical differences between the CE and MCE treatments of BH decay in the case of 
the EH action by setting $\alpha=\beta=\gamma=0$; the results are shown in 
Fig.~\ref{fig3}. The first thing to examine is the mass loss rate of BH; to this 
end we consider the dimensionless quantity $M_*^{-2}dM/dt=dm/dt$, with time measured in 
Planck units, which is shown in the upper panel in the Figure 
for $n=3,5$ for both the CE and MCE cases. For larger values of $m$, 
both the MCE and CE analyses yield identical results as expected but differ 
significantly once $m \lsim 4$. In the CE case, the rates grow rapidly as the mass 
of the BH decreases below this range whereas in the MCE case the rate drops quite 
dramatically. The influence on the lifetimes of these significantly different 
behaviors is shown in the lower panel for a BH which begins life with $m=5$, a 
reasonable value for production at the LHC. Here what is specifically 
shown is the time taken(in units of $M_*^{-1}$) for an initial $m=5$ BH to decay 
to a BH with $m<5$ via Hawking radiation for various values of $n$. Note that 
while the state of complete BH evaporation, $m=0$, is reached in the CE case for 
$0.03 \lsim M_*t \lsim 1$, for the MCE analysis one obtains values which are larger 
than this by many(6 to 8 or more) orders of magnitude. This shows, assuming the 
EH action, that the slowing of the 
BH evaporation rate in the MCE approach leads to a dramatic increase in the lifetime 
of BH that can be produced in TeV collisions as has been emphasized by several 
sets of authors{\cite {Casadio:2001dc}}.  

\begin{figure}[htbp]
\centerline{
\includegraphics[width=8.5cm,angle=90]{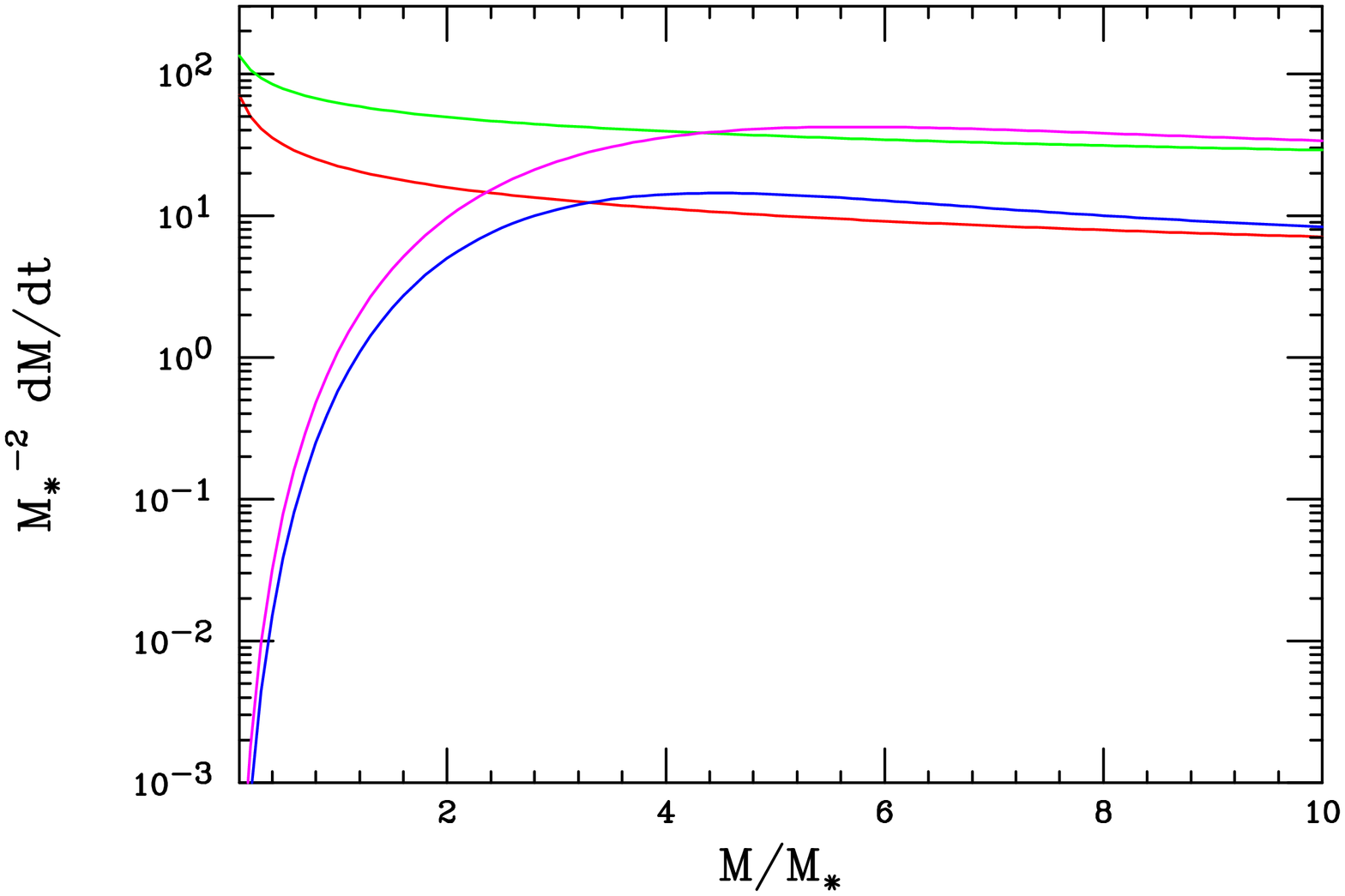}}
\vspace{0.4cm}
\centerline{
\includegraphics[width=8.5cm,angle=90]{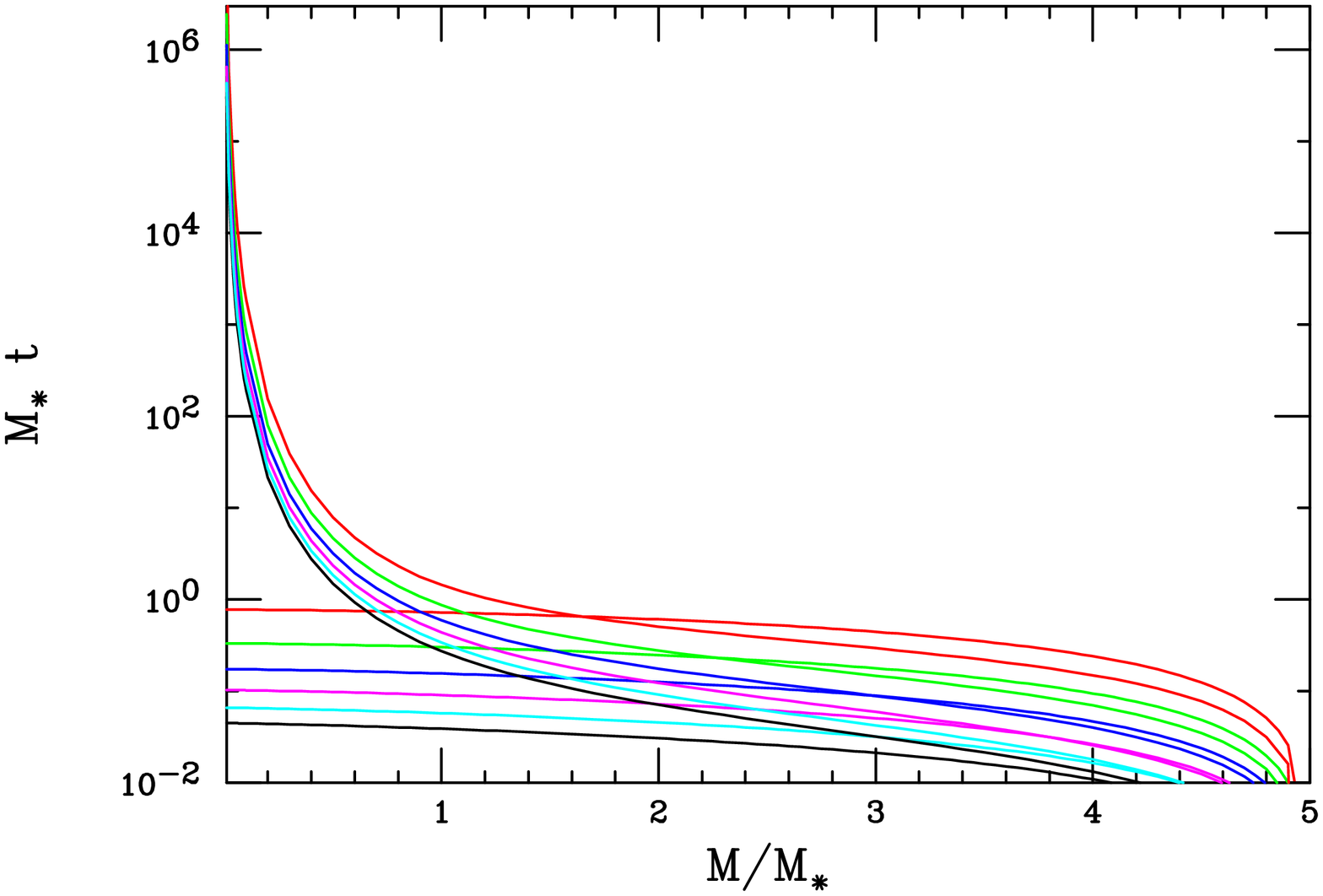}}
\vspace*{0.1cm}
\caption{(Top) Rate of change of the BH mass through Hawking radiation on the brane 
assuming the EH action as discussed in the text; the CE(top) and MCE(bottom) 
results are represented by the two sets of curves with the case $n=5(3)$ being 
the upper(lower) curve. (Bottom) Decay times for a BH with an initial $m=M_{BH}/M_*=5$; 
the rising(flat) set of curves at low $m$ corresponds 
to the MCE(CE) case. In each set of curves $n$ ranges from 2 to 7 going from top to bottom.}
\label{fig3}
\end{figure}

We now turn to the case where Lovelock invariants are present in the action. In order 
to analyze a scenario where the number of extra dimensions is even, 
we consider a specific example with $n=2$ so that only the quadratic 
G-B term can be present, \ie, only $\alpha$ can be non-zero. Note that as 
$n$ is even, the G-B invariant cannot lead to a BH mass threshold.  
Fig.~\ref{fig4} shows the results of the calculations 
which are analogous to those displayed in the previous Figure. 
Here in the upper panel we see that for both 
the MCE and CE analyses the presence of the G-B term leads to a suppression 
of the BH decay rate, $M_*^{-2}dM/dt$, which is active for all values of $m$ but is 
somewhat magnified as $m$ gets smaller. For larger $m$ the CE and MCE analyses are seen 
to agree just as in the case of the EH action above. For the overall BH lifetime,  
$M_*t$, we see in the bottom panel that a non-zero $\alpha$ for the case of $n=2$ can 
increase the value of this quantity up to 2 orders of 
magnitude in the CE analysis. Using the 
MCE combined with the non-zero values of $\alpha$ leads to a further  
increase in the BH lifetime by two more orders of magnitude, \ie, the presence of the G-B  
term is seen to further augment the BH lifetime obtained by employing the MCE 
analysis in comparison to the standard EH picture employing the CE. However, this 
relative enhancement is far smaller than that found in the EH case since the common 
enhancement arising from the Lovelock terms present in both cases is already large. 
We thus conclude that for $n$ even the presence of Lovelock invariants together with 
the use of the MCE will significantly increase the BH lifetimes but only by a few  
orders of magnitude. 
\begin{figure}[htbp]
\centerline{
\includegraphics[width=8.5cm,angle=90]{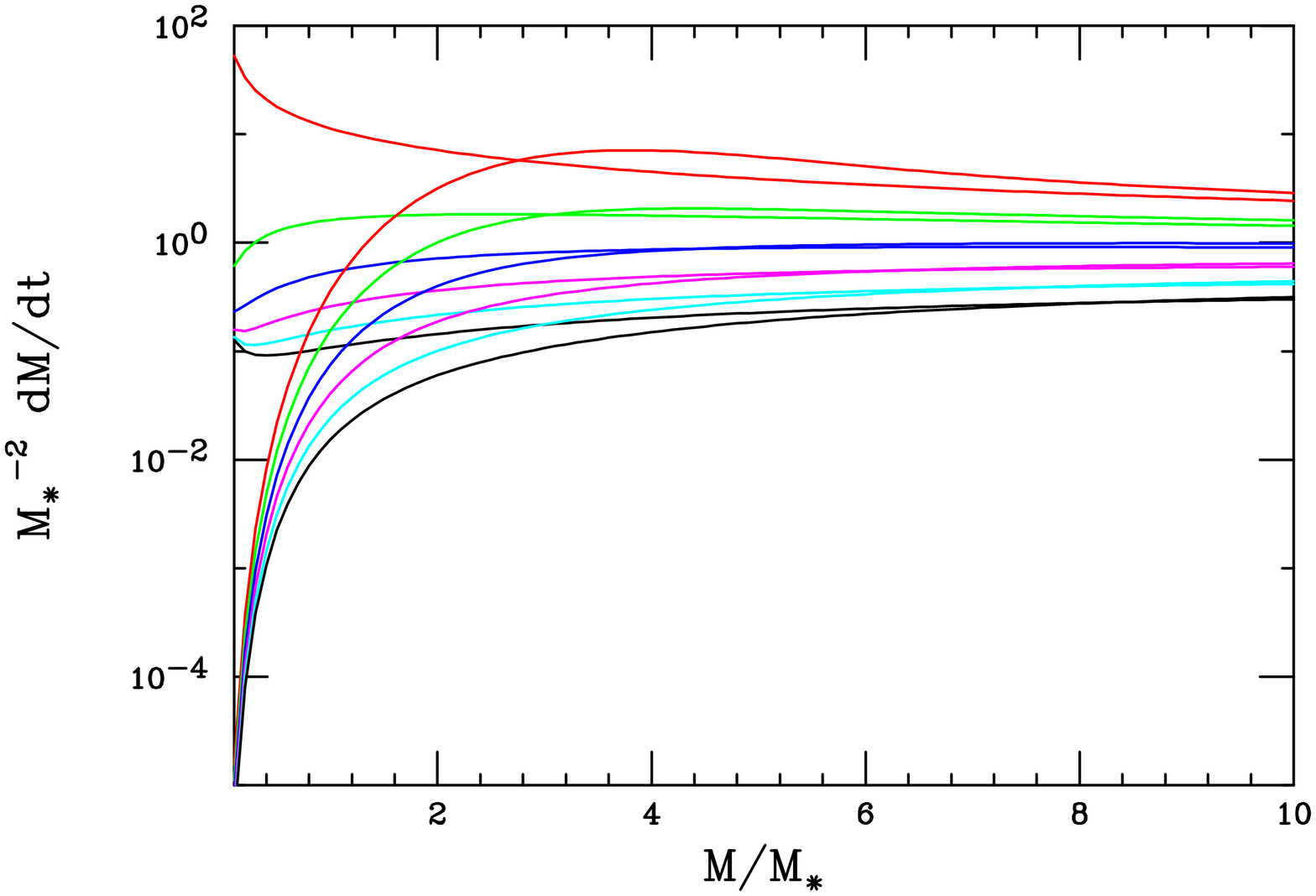}}
\vspace{0.4cm}
\centerline{
\includegraphics[width=8.5cm,angle=90]{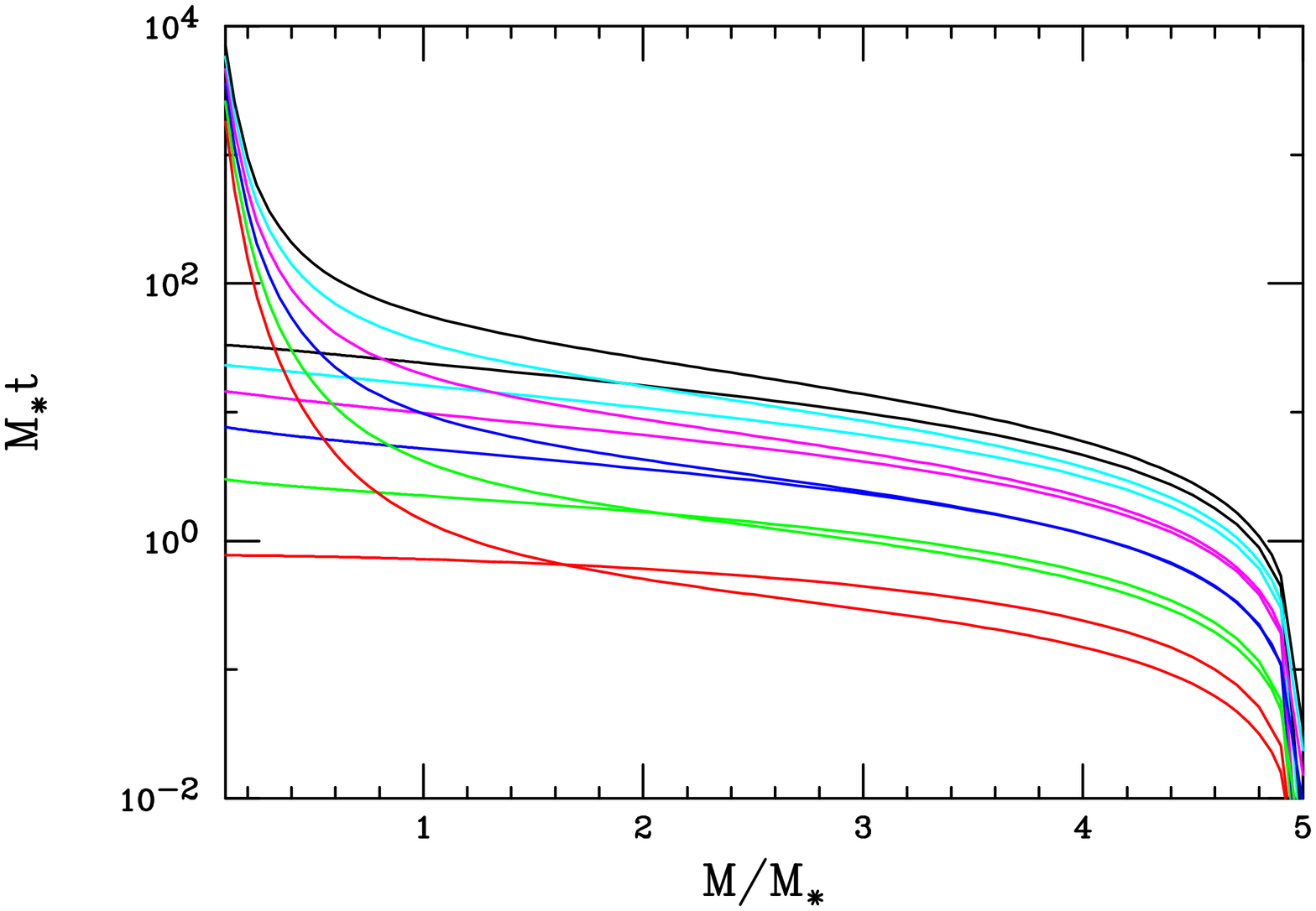}}
\vspace*{0.1cm}
\caption{Same as the previous figure but now assuming $n=2$ for $\alpha \neq 0$. In 
the top panel, for both sets of curves, $\alpha$ goes from 0 to 0.025 in steps of 0.005 
going from top to bottom. In the bottom panel, $\alpha$ goes over the same range but in 
opposite order.}
\label{fig4}
\end{figure}

What happens when $n$ is odd and a BH remnant can form? In this case we consider typical  
non-zero values for all of the allowed Lovelock parameters for the fixed number of extra 
dimensions considered. Figs.~\ref{fig5} and ~\ref{fig6} 
show the values of  $M_*^{-2}dM/dt$ for $n$=3 and 5 comparing the expectations of the 
CE and MCE approaches. For the case of $n=3$, we hold $\beta=0.0005$ fixed and 
vary the values of $\alpha$; for $n=5$, we hold fixed $\gamma=1.14\times 10^{-5}$ and 
vary the values of both $\alpha$ and $\beta$. In both figures we see that the general 
patterns associated with a reduced rate of Hawking radiation employing the MCE analysis 
observed for $n=2$ repeated. However, unlike for even values of $n$, for $n$ odd we see 
that $M_*^{-2}dM/dt \to 0$ for {\it both} the MCE and CE approaches as $m \to m_{crit}$. 
In addition, for large values of $m$, 
$M_*^{-2}dM/dt$ is generally seen to be larger when the CE technique is employed 
than when one instead uses the MCE.

\begin{figure}[htbp]
\centerline{
\includegraphics[width=8.5cm,angle=90]{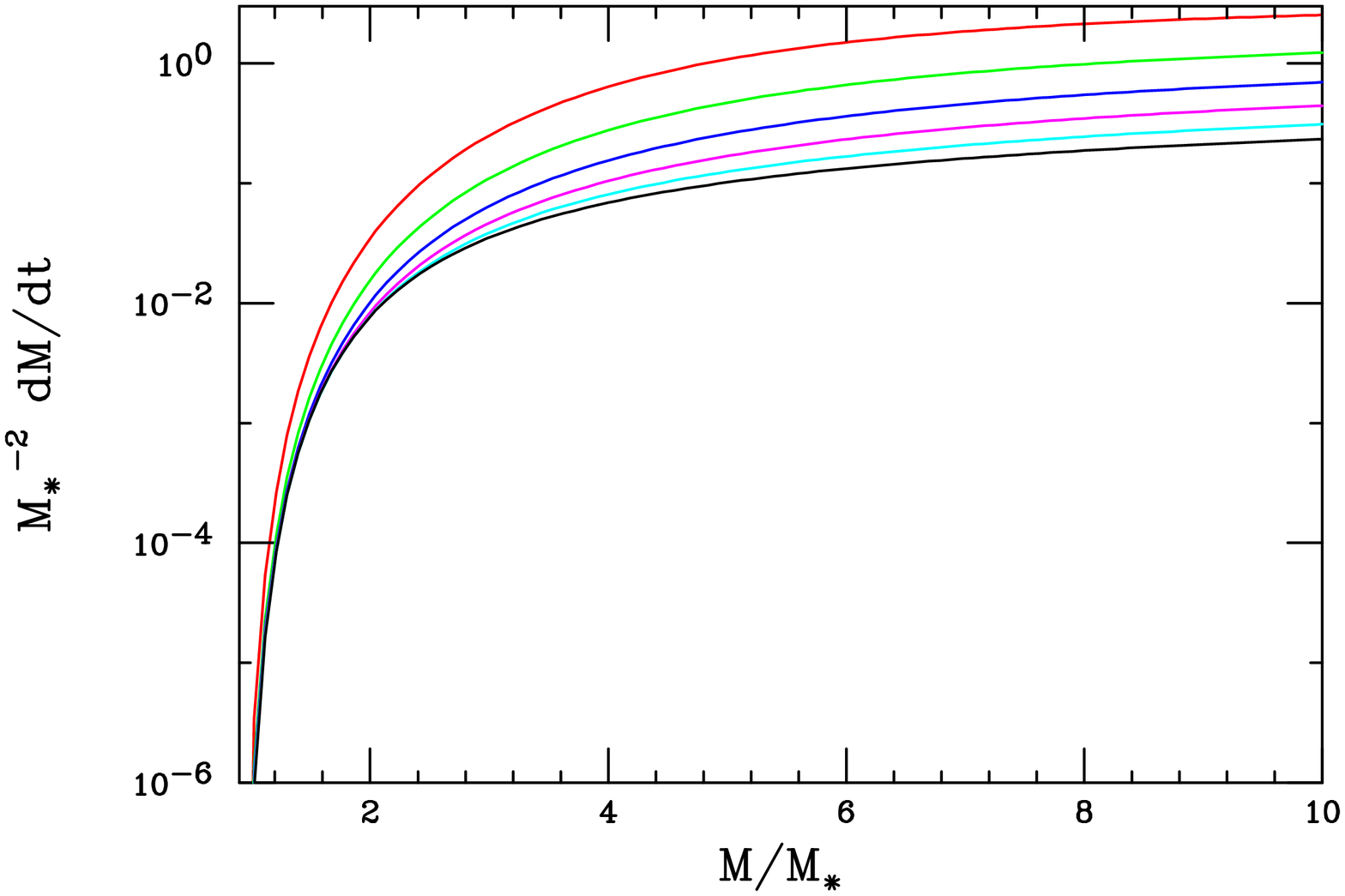}}
\vspace{0.4cm}
\centerline{
\includegraphics[width=8.5cm,angle=90]{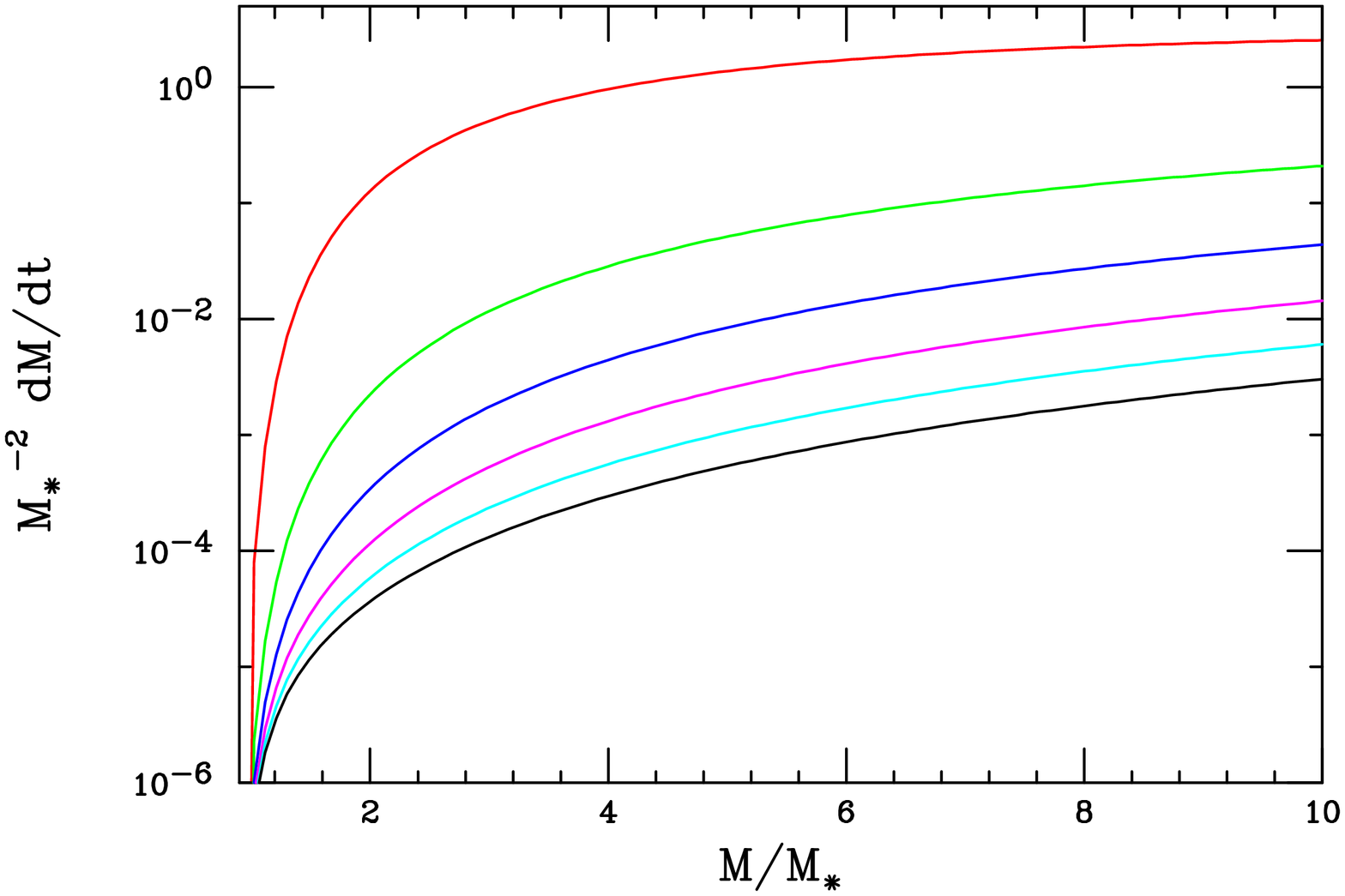}}
\vspace*{0.1cm}
\caption{BH decay rate as a function of $m=M/M_*$ for $n=3$ with $\beta=0.0005$. From 
top to bottom the value of $\alpha$ ranges from 0 to 0.025 in steps of 0.005. The 
upper(lower) panel employs the CE(MCE) analysis approach.}
\label{fig5}
\end{figure}

In order to access the influence on the BH decay time of the CE versus MCE choice in the 
presence of higher curvature terms, we show in Figs.~\ref{fig7} and ~\ref{fig8} the 
integrated decay time for the above studied cases of 
$n=3$ and 5. As expected from the $n=2$ analysis, here we again see that the BH described 
by the MCE has a decay time for $m> m_{crit}$ which is somewhat longer than the 
corresponding CE result. This enhancement 
in the decay time to a fixed mass final state, which can be up to a couple of orders of 
magnitude, is observed to be substantially less than that 
obtained in the pure EH scenario found above. We thus conclude, for $n$ odd, that 
while the MCE 
approach always leads to an enhancement in the BH decay time relative to that obtained 
in the CE approach, the effect of the higher Lovelock terms is to reduce the degree of this 
enhancement in comparison to that obtained in the case of 
the pure EH action. Note that in {\it either} approach the resulting {\it total} BH 
lifetime is infinite as the BH decay still results in a 
stable remnant as was found in our earlier work.

\begin{figure}[htbp]
\centerline{
\includegraphics[width=8.5cm,angle=90]{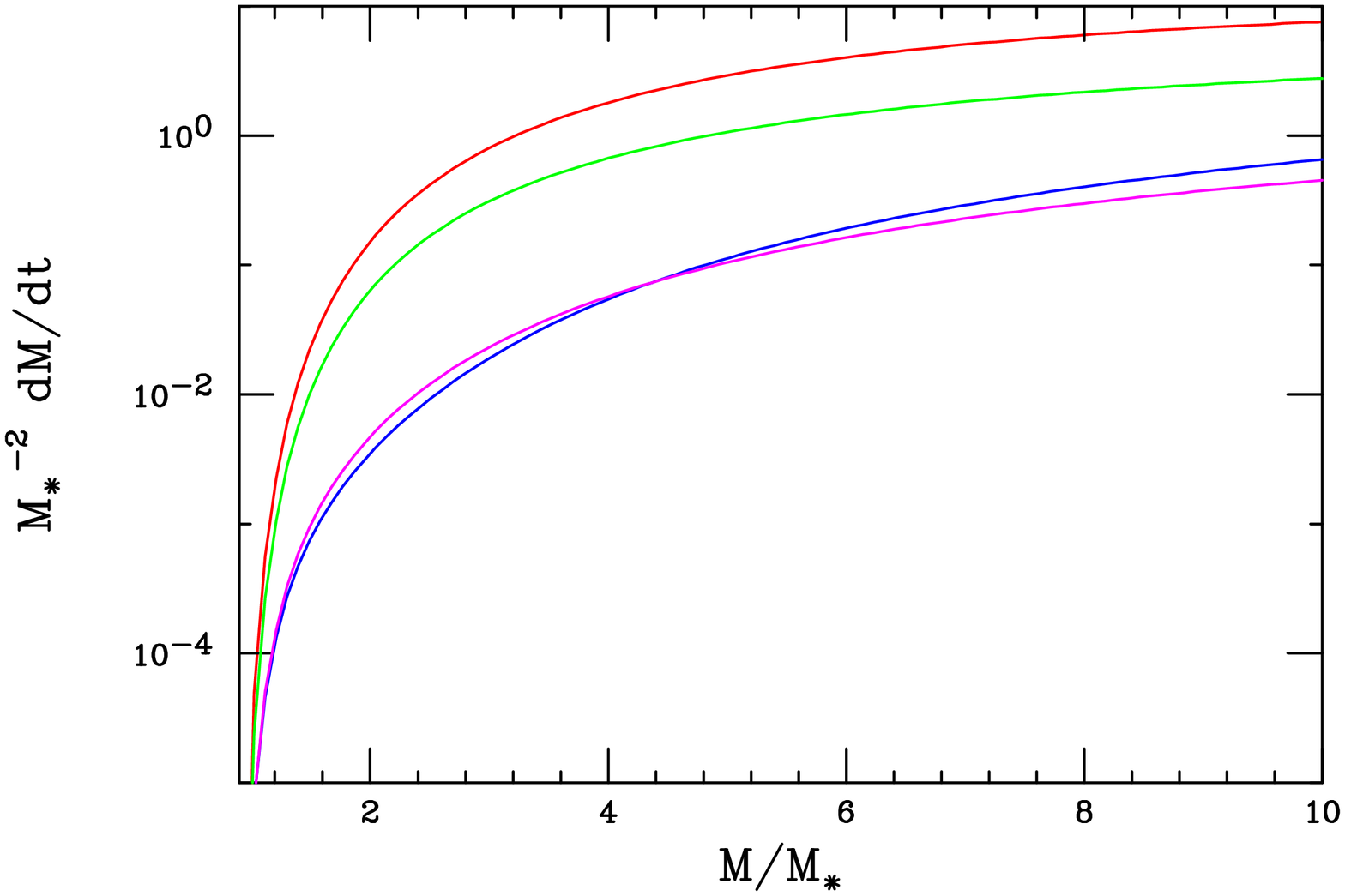}}
\vspace{0.4cm}
\centerline{
\includegraphics[width=8.5cm,angle=90]{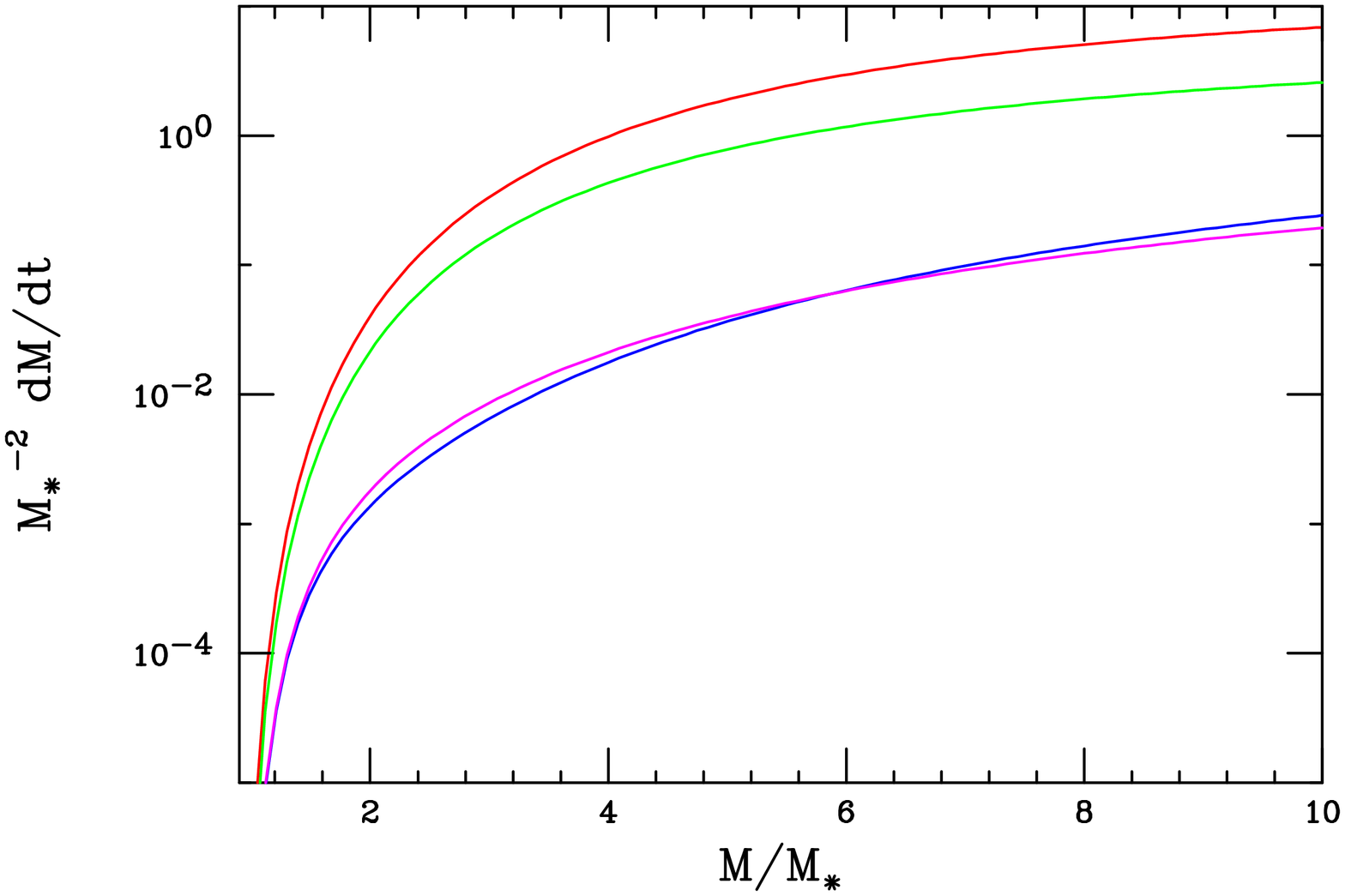}}
\vspace*{0.1cm}
\caption{Same as the previous figure but now for $n=5$ with $\gamma=1.14\times 10^{-5}$. 
From top to bottom on the right-hand side of the plot the curves correspond to 
$\alpha=\beta=0$, $\alpha=0.005$, $\beta=0.0003$ and $\alpha=0.005,~\beta=0.0003$, 
respectively.}
\label{fig6}
\end{figure}

It is clear from this analysis that the Lovelock extended action leads to significant 
modifications in the mass loss rates and lifetimes of BH and that the choice of MCE vs CE 
is critical. Theoretical arguments support the use of the MCE description but experiments 
will be able to distinguish these two approaches at colliders.

\begin{figure}[htbp]
\centerline{
\includegraphics[width=8.5cm,angle=90]{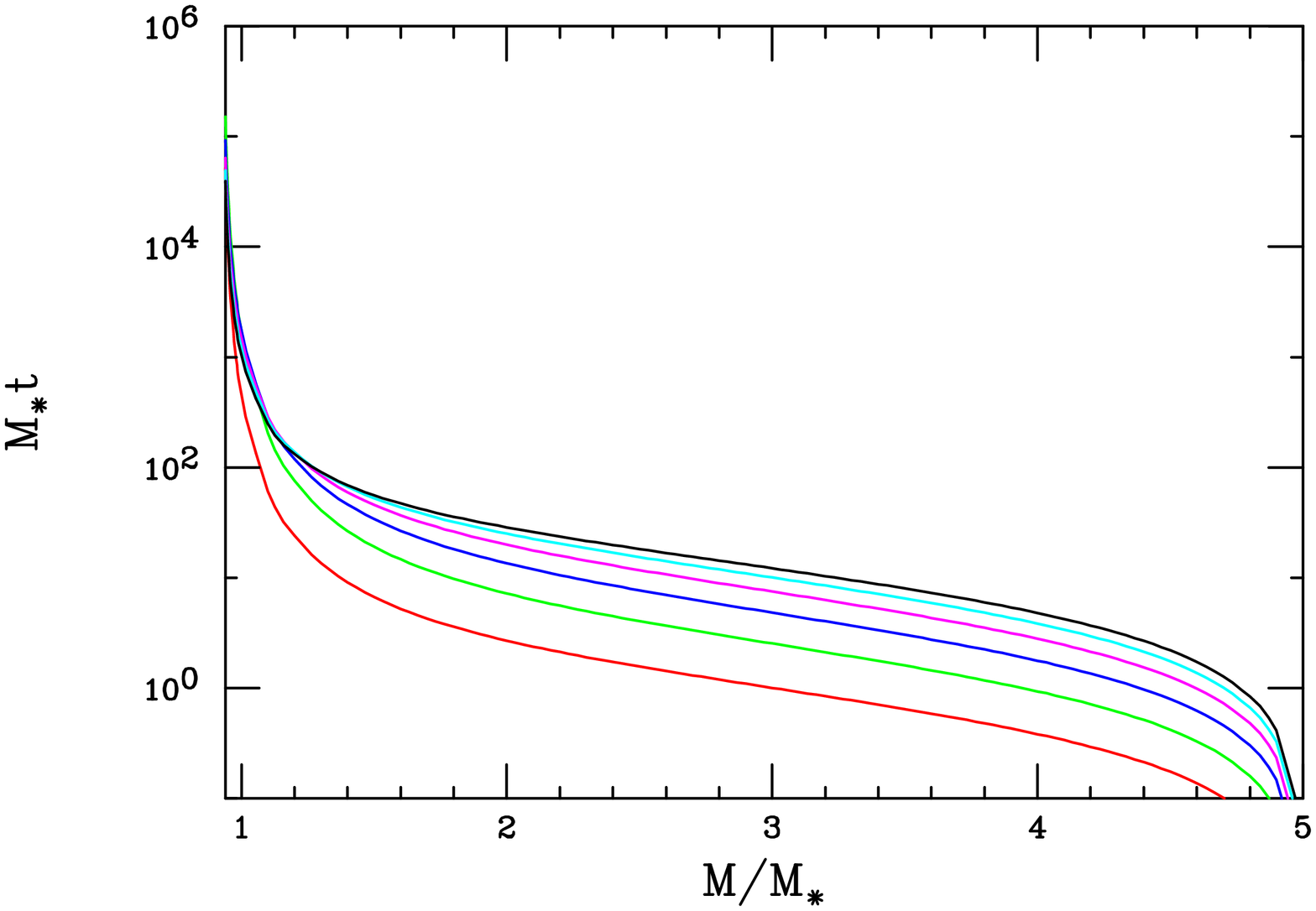}}
\vspace{0.4cm}
\centerline{
\includegraphics[width=8.5cm,angle=90]{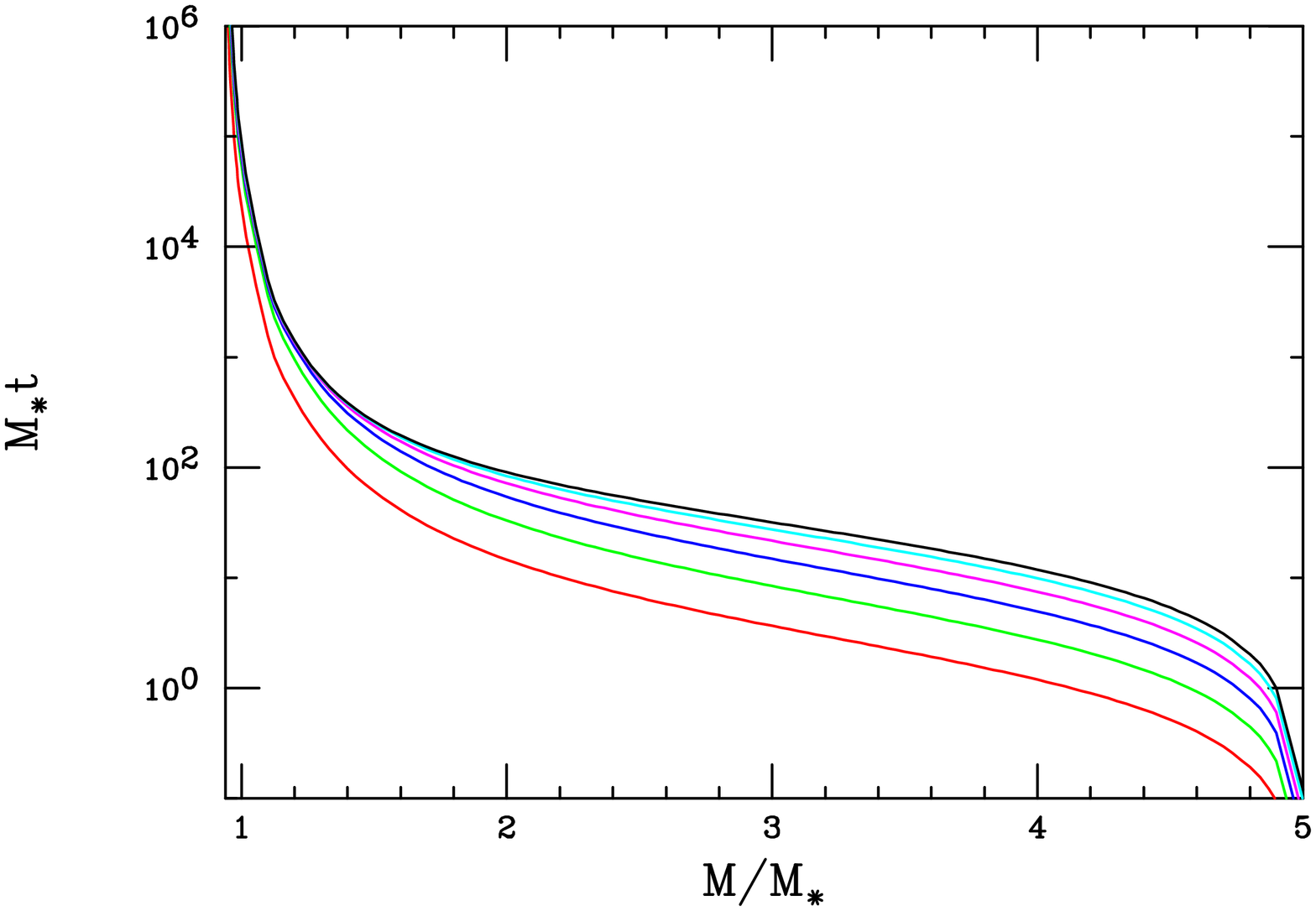}}
\vspace*{0.1cm}
\caption{Integrated BH lifetimes corresponding to the decay rates shown in Fig.5 
but with the curves labeled in the opposite order.}
\label{fig7}
\end{figure}

\section{Discussion and Conclusions}

Higher curvature invariants of the Lovelock type could be present in the extra-dimensional 
effective gravitational action and would make their presence known 
at energies of order $M_*$ and above. If 
the higher dimensional bulk is flat (as in ADD-like models but not in RS-like models) there 
are few ways to directly probe the existence of these additional terms experimentally. The 
reason for this is that the conventional `perturbative' graviton-related ADD signatures are 
found to be quite insensitive to the existence of Lovelock terms. On the otherhand, 
since they are non-perturbative structures, the properties of TeV-scale black holes in 
extra-dimensional models are potentially very sensitive to these new interactions which   
can be probed at future particle colliders.

\begin{figure}[htbp]
\centerline{
\includegraphics[width=8.5cm,angle=90]{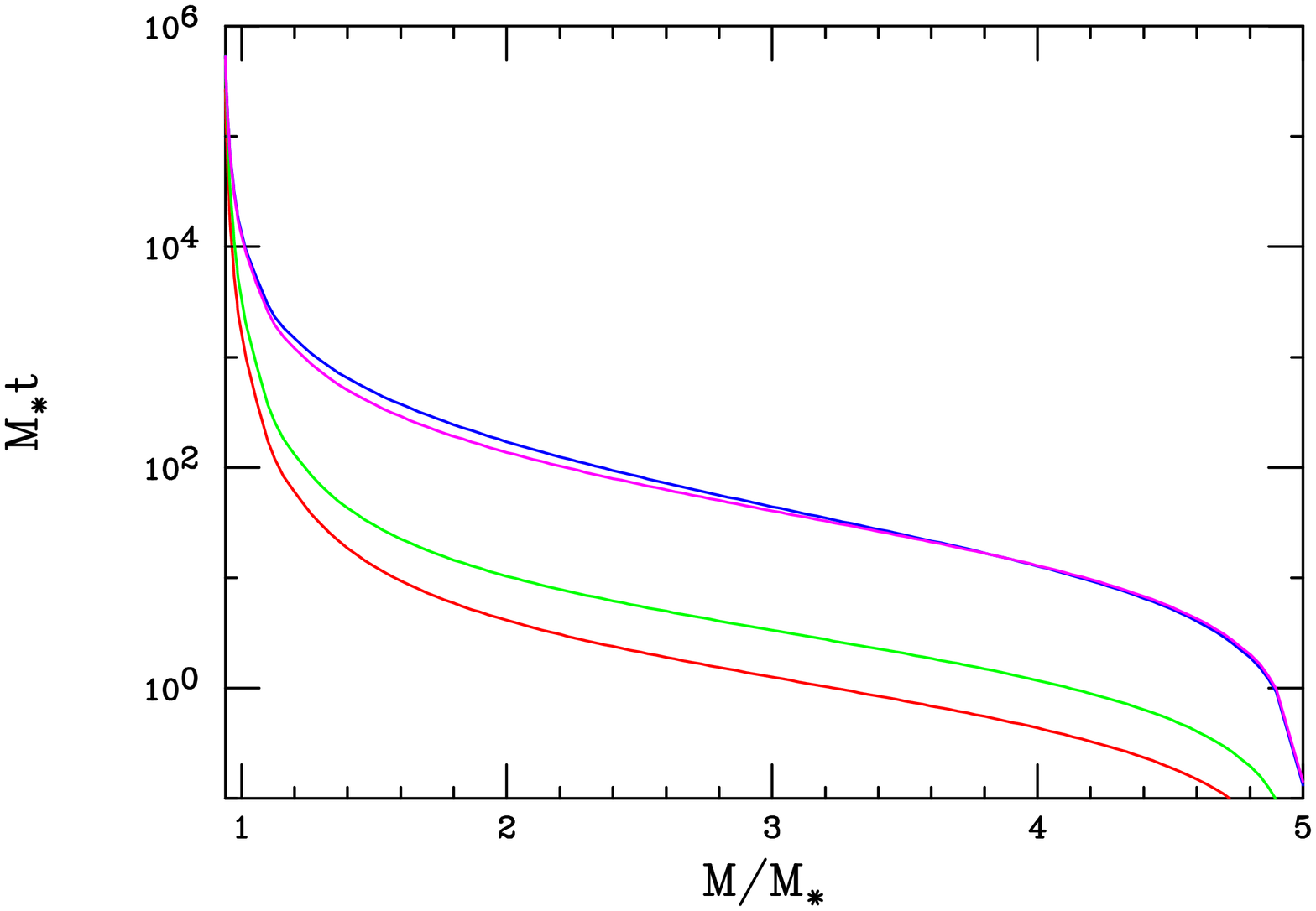}}
\vspace{0.4cm}
\centerline{
\includegraphics[width=8.5cm,angle=90]{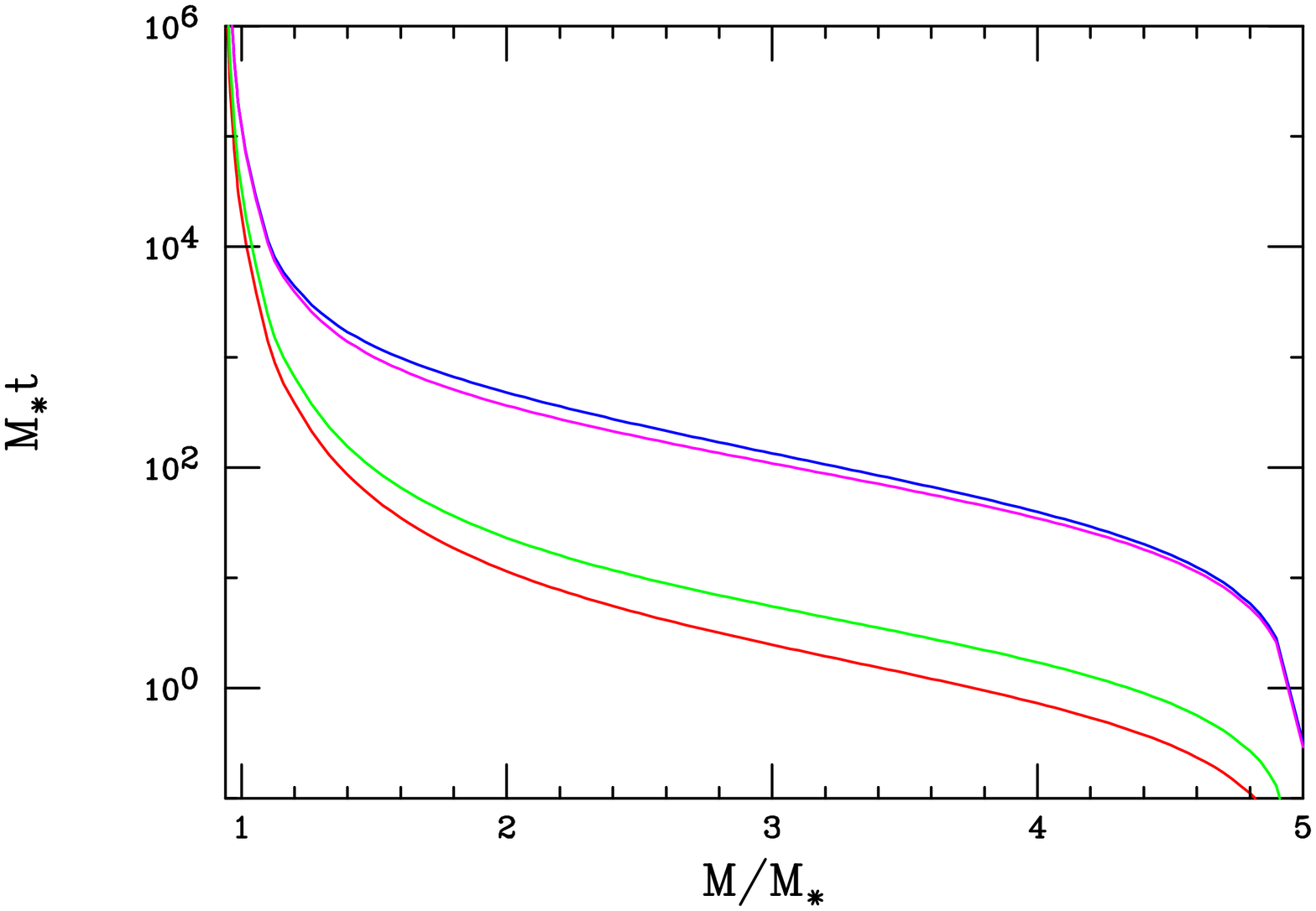}}
\vspace*{0.1cm}
\caption{Integrated BH lifetimes corresponding to the decay rates shown in Fig.6 
but with the curves labeled in the opposite order.}
\label{fig8}
\end{figure}

In this paper we have considered how the existence of Lovelock invariant extensions to 
the Einstein-Hilbert action will modify the mass loss rates and lifetimes of TeV-scale BH. In 
particular we examined the sensitivity of both of these quantities to the choice made in 
the statistical mechanics treatment of BH. It had been shown, and we verified those results 
here, that in the case of the EH action, BH lifetimes are significantly enhanced by many 
orders of magnitude when the microcanonical ensemble description is employed in comparison 
to the more conventional canonical ensemble approach. There are several reasonable arguments 
in the literature as to why BH in the mass range of interest to us are in fact best described 
by the MCE. 

Within this context, when Lovelock 
terms are present in the case of ADD-like flat extra dimensions we demonstrated in the 
present paper that: ($i$) BH decays to SM fields on the brane remain dominant over those 
to graviton bulk fields employing either the MCE or CE descriptions when Lovelock terms 
are present. However, in all cases the bulk/brane ratio was shown to grow as the number of 
extra dimensions increases. ($ii$) Unlike in the case of a single warped extra dimension, 
the BH 
decay rates and lifetimes for ADD-like extra dimensions are found to be insensitive to the 
statistics `mix'  
of the particles on the brane. ($iii$) For {\it even} numbers of extra dimensions the 
lifetimes of BH described by the MCE are up to a few orders of magnitude larger than those  
obtained employing the CE. While this significant enhancement is large it is many orders 
of magnitude smaller than that obtained employing only the EH action. ($iv$) For odd numbers 
of extra dimensions, with the highest order allowed Lovelock term present, BH are found 
to decay to stable relics independent of the MCE/CE choice. However, the functional 
dependence of the mass loss rate in the two cases can be somewhat different but the details 
are sensitive to the particular values of the model parameters.  
It is interesting to note that the existence of a remnant and a BH mass threshold 
in models with Lovelock invariants in 
the action is not an uncommon feature of models which probe beyond the EH action: such phenomena 
may happen for a 4-d BH when a renormalization group running 
of Newton's constant is employed{\cite {Bonanno}} in order to approximate 
leading quantum corrections. Such a remnant scenario can also be seen to 
occur in theories with a minimum length{\cite {cav}}, in loop quantum 
gravity{\cite {loop}} and in resummed quantum gravity{\cite {Physics:2006vw}}. In, \eg, 
the case of a minimum length, stable remnants occur for all numbers of extra dimensions. 
It is interesting to note that this 
phenomena occurs in all these models where one tries to incorporate 
quantum corrections in some way; though the quantitative nature of such remnants differ 
in detail in each of these models, it would be interesting to learn whether or 
not this is a general qualitative feature of all such approaches. 

Black holes observed at future colliders may open an exciting window on the fundamental 
theory of gravity in extra dimensions.

\noindent{\Large\bf Acknowledgments}

The author would like to thank J.Hewett and B. Lillie for discussions 
related to this work.

%
\def\MPL #1 #2 #3 {Mod. Phys. Lett. {\bf#1},\ #2 (#3)}
\def\NPB #1 #2 #3 {Nucl. Phys. {\bf#1},\ #2 (#3)}
\def\PLB #1 #2 #3 {Phys. Lett. {\bf#1},\ #2 (#3)}
\def\PR #1 #2 #3 {Phys. Rep. {\bf#1},\ #2 (#3)}
\def\PRD #1 #2 #3 {Phys. Rev. {\bf#1},\ #2 (#3)}
\def\PRL #1 #2 #3 {Phys. Rev. Lett. {\bf#1},\ #2 (#3)}
\def\RMP #1 #2 #3 {Rev. Mod. Phys. {\bf#1},\ #2 (#3)}
\def\NIM #1 #2 #3 {Nuc. Inst. Meth. {\bf#1},\ #2 (#3)}
\def\ZPC #1 #2 #3 {Z. Phys. {\bf#1},\ #2 (#3)}
\def\EJPC #1 #2 #3 {E. Phys. J. {\bf#1},\ #2 (#3)}
\def\IJMP #1 #2 #3 {Int. J. Mod. Phys. {\bf#1},\ #2 (#3)}
\def\JHEP #1 #2 #3 {J. High En. Phys. {\bf#1},\ #2 (#3)}

\end{document}